\documentclass[aps,prb,twocolumn,groupedaddress]{revtex4}

\usepackage{graphicx}
\usepackage{amsmath,amsfonts,amssymb,bbm}

\begin{document}

\title{Statistics of radiation emitted from a quantum point contact}

\author{A.V.\ Lebedev$^{\, a,b}$, G.B.\ Lesovik$^{\, b}$,
and G.\ Blatter$^{\, a}$}

\affiliation{$^{a}$Theoretische Physik, ETH-H\"onggerberg, CH-8093
Z\"urich, Switzerland}

\affiliation{$^{b}$L.D.\ Landau Institute for Theoretical Physics, RAS,
117940 Moscow, Russia}

\date{\today}

\begin{abstract}

We analyze the statistics of the electromagnetic radiation emitted from
electrons pushed through a quantum point contact. We consider a setup
implemented in a two-dimenional electron gas (2DEG) where the radiation
manifests itself in terms of 2D plasmons emitted from electrons scattered at
the point contact.  The bosonic statistics of the plasmons competes with the
fermionic statistics of the electrons; as a result, the quantum point contact
emits non-classical radiation with a statistics which can be tuned from
bunching to anti-bunching by changing the driving voltage. Our perturbative
calculation of the irreducible two-plasmon probability correlator provides us
with information on the statistical nature of the emitted plasmons and on the
underlying electronic current flow.

\end{abstract}

\maketitle

\section{Introduction}

The interest in photon radiation from a quantum point contact (QPC) is
two-fold: on the one hand, the quantum point contact acts as a source of
non-classical light \cite{beenakker_01,beenakker_04}. While a classical
current produces photons with Poisson statistics \cite{glauber_63}, the
current pushed through a QPC can be tuned to radiate photons with either
super- or sub-poissonian statistics \cite{beenakker_01,beenakker_04},
testifying for the bosonic nature of the photons (bunching) or the fermionic
statistics (anti-bunching) of the underlying electrons radiating these
photons, e.g., as `Bremsstrahlung' radiation in back-reflection processes.  On
the other hand, the radiation statistics contains information on the
statistics of current fluctuations across the QPC---as such, the photodetector
serves as a tool to probe the nearby current flow in the mesoscopic wire
\cite{lesoviklosen_97,gavish_00}.  Both themes have attracted considerable
interest, be it in the context of the counting statistics of electrons in
phase coherent mesoscopic conductors \cite{levitov_96,nazarov, ensslin_fcs},
where higher-order correlators carry the signatures of interactions, or be it
related to the search for new sources of non-classical radiation
\cite{beenakker_01,beenakker_04}. 

The classic photodetection theory for optical photons goes back to Glauber
\cite{glauber_63} and is based on a threshold detector, where each
(sufficiently energized) photon induces an electronic cascade generating a
counting signal. The theory has been widely applied in the optical regime
where thres\-hold energies reside in the $e$V regime.  The analysis of the
photon counting statistics in Refs.\ \onlinecite{beenakker_01,beenakker_04}
has been based on this photodetection theory and has concentrated on the
regime, where the energy of emitted photons is larger then $eV/2$, with $V$
the voltage bias applied to the quantum point contact. For this situation, the
emitted photons are anti-bunched, since they are produced by different
electrons scattered by a QPC.

The application of Glauber's theory to mesoscopic setups is not
straightforward, however, as typical photons generated in a mesoscopic
structure have energies in the GHz regime and it may be difficult to construct
a stable photodetector with a correspondingly small threshold energy.
Furthermore, the emission rate of free photons scales with the small factor
$\alpha (v_{\rm\scriptscriptstyle F}/c)^2 \sim 10^{-8}$, where $\alpha$ is the
fine structure constant, $v_{\rm\scriptscriptstyle F}$ is the Fermi velocity,
and $c$ the velocity of light, rendering the observation of free photons
practically impossible.
\begin{figure} [t]
   \includegraphics[width=8cm]{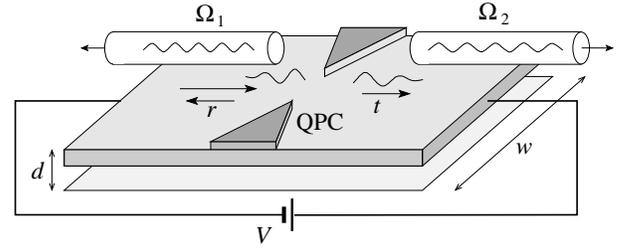}
   \caption[]
   {Quantum point contact (QPC) fabricated through gates constricting the
   electron flow in a two-dimensional electron gas (2DEG). We consider a
   quasi-1D setup where the width $w$ of the 2DEG allows propagation of one
   transverse mode; $d$ is the distance to the backgate.  The voltage ($V$)
   driven quantum point contact emits 2D plasmons with frequencies $\Omega_1$
   and $\Omega_2$ due to electron backscattering ($t$ and $r$ denote
   transmission and reflection amplitudes). These plasmons are picked up by
   waveguides transmitting the signal to the measurement setup.  The
   measurement of the one- [$P_1(\Omega)$] and two-plasmon [$P_2(\Omega_1,
   \Omega_2)$] emission probabilities allows to characterize the radiation and
   provides information on the fourth-order current correlator.}
   \label{fig:setup}
\end{figure}

On the other hand, within a mesoscopic context, the low-energy exchange
between a quantum point contact and the electromagnetic environment can be
studied with a double-dot detector\cite{aguado_00,Kueng_09}.  Furthermore, a
setup has been analyzed recently, where low-energy radiation manifests itself
in terms of plasmons propagating within the two-dimensional electron gas
(2DEG) forming the quantum point contact \cite{gabelli_04,portier}, cf.\ Fig.\
\ref{fig:setup}. These 2D plasmons then are further transmitted through
waveguides and then analyzed in a power detector.  The emission rate for
plasmons is enhanced over that of free photons by a factor
$(c/v_\mathrm{pl})^3 \sim 10^6$, where $v_\mathrm{pl}$ denotes the plasmon
velocity; this gain in signal has to be preserved by proper impedance matching
of the waveguides and the quantum point contact.  In this paper, we study this
kind of setup and analyze the two-plasmon correlations in an arbitrary
frequency range; these correlations then tell about the statistics of the
emitted radiation (bunching or anti-bunching) and carry information on the
fourth-order current correlator in the mesoscopic wire.

Below, we present a perturbative calculation (in the QPC--electromagnetic
field coupling) of the one- [$P_1(\Omega)$] and two-plasmon
[$P_2(\Omega_1,\Omega_2)$] probabilities for emission at given frequencies
$\Omega,~\Omega_{1,2}$ and during an arbitrary but fixed time $t_0$.  Within
our perturbative approach, we calculate the irreducible probability correlator
$\bar{P}_2(\Omega_1,\Omega_2) = P_2(\Omega_1,\Omega_2) - P_1(\Omega_1)
P_1(\Omega_2)$.  The sign of this quantity then tells us about the character
of the emitted radiation, bunching or anti-bunching. E.g., a positive sign
indicates that plasmons prefer to be emitted simultaneously rather then
independently and thus are bunched, while for a negative sign the opposite
situation of anti-bunching prevails; for $\bar{P}_2(\Omega_1, \Omega_2)=0$ the
two plasmons are emitted independently. As we will show below, changing the
bias on the quantum point contact will allow to change the statistics from
anti-bunched fermionic-type to bunched bosonic-type.  Furthermore, for
different frequencies $\Omega_1 \neq \Omega_2$, the quantity
$\bar{P}_2(\Omega_1,\Omega_2)$ is directly proportional to the irreducible
fourth-order current correlator and hence its measurement provides valuable
insight into the fluctuation statistics of the current across the quantum
point contact.

In the following, we first define the model and find the expressions for the
probability densities $p_1(\Omega)$ and $p_2(\Omega_1,\Omega_2)$ expressed
through current correlators within a perturbative expansion, see Sec.\
\ref{sec:prob}.  In a second step, these probability densities are rewritten
through the second and fourth order irreducible noise correlators $S^{(2)}$
and $S^{(4)}$. In section \ref{sec:count}, we combine results for the
probability densities and the noise correlators to find the single photon
emission probability $P_1(\Omega) = p_1(\Omega) d\Omega$, the correlated
two-photon emission probability $P_2(\Omega_1,\Omega_2)=p_2(\Omega_1,\Omega_2)
d\Omega_1 d\Omega_2$, as well as the irreducible quantity $\bar{P}_2
(\Omega_1,\Omega_2)$; we identify the regimes of anti-bunching at intermediate
voltages and the competition between anti-bunching and bunching at high
voltage, when one electron has sufficient energy to emit two (bunched)
plasmons. In section \ref{sec:current}, we discuss the interrelation between the
fourth-order current correlator and the irreducible correlator $\bar{P}_2
(\Omega_1,\Omega_2)$ and we conclude in Sec.\ \ref{sec:conclusion}.

\section{Photon emission probabilities}
\label{sec:prob}

In this section, we derive the formal expressions for the probability
densities $p_1(\Omega)$ and $p_2(\Omega_1,\Omega_2)$ to emit one or two
plasmons at frequencies $\Omega,~\Omega_{1,2}$ during a given time and express
the results through the second- and fourth-order current correlators. We
restrict ourselves to a perturbative analysis in the Hamiltonian $\hat
H_\mathrm{int}$, describing the interaction between the electronic current
density $\hat{\bf j}({\bf r})$ through the QPC and the electromagnetic field
described through the vector potential $\hat {\bf A}({\bf r})$,
\begin{equation}
      \hat H_\mathrm{int} = -\frac1c \int d^3r \, \hat{\bf j}({\bf r})
      \cdot \hat {\bf A}({\bf r}).
      \label{hamAj}
\end{equation}
In a realistic setup, the electromagnetic field is generated by the
two-dimensional plasmon modes of the two-dimensional electron gas
(2DEG)~\cite{burke}. These plasmon modes propagate along the one-dimensional
wire (the $x$ direction) and we consider the lowest transverse mode only (we
assume translation invariance along the $y$ direction). The presence of the
backgate changes the dispersion from the usual (in 2D) square-root form to a
linear one.  The vector potential $\hat {\bf A}(x)$ then has only a component
along $x$,
\begin{equation}
      \hat{A}_x(x) = \sum_{k}
      ik \gamma\Bigl(\frac{2\pi\hbar c^2}{\omega_k L_{\rm quant}}\Bigr)^{1/2}
      \bigl( \hat c_{k} \, e^{ikx}
      - \hat c_{k}^\dagger\, e^{-ikx}\bigr),
\end{equation}
where $\hat c_{k}^\dagger$ ($\hat c_{k}$) are bosonic creation (annihilation)
operators for the (longitudinal) plasmon modes with wave vector $k$; $\omega_k
= v_\mathrm{pl} k$ is the plasmon frequency with $v_\mathrm{pl}$ the plasmon
wave velocity and $\gamma\sim\sqrt{d/w}$ is a geometrical factor with $w$ the
width of the 2DEG and $d$ the distance between the 2DEG and a Schottky barrier
electrostatic gate\cite{burke}. In the following, we set the quantization
length $L_{\rm quant}$ equal to unity.

With the current operator $\hat{j}_x({\bf r})=\hat{I}(x)
\delta^3({\bf r}-x {\bf e}_x)$ in the wire, the Hamiltonian
$\hat{H}_\mathrm{int}$ in the interaction representation takes the
form
\begin{eqnarray}
      \label{eq:H_int}
      &&\hat{H}_\mathrm{int}(t) = -i\gamma\sum_{k} k\Bigl(
      \frac{2\pi\hbar}{\omega_k} \Bigr)^{1/2}\!\!
      \\
      &&\nonumber
      \qquad\times \Bigl( \hat{c}_k\hat{I}_k(t)
      e^{-i\omega_k t} - \hat{c}_k^\dagger
      \hat{I}_k^\dagger(t) e^{i\omega_k t}\Bigr),
\end{eqnarray}
where $\hat I_k(t)$ is the spatial average of the current $\hat I(x,t)$ over
the coupling region of the plasmon described by the kernel $f(x)$ with
extension $L$, 
\begin{equation}
      \hat I_k(t) = \int dx\, \hat I(x,t)\,f(x)\,e^{ikx}.
      \label{Ik}
\end{equation}
In a typical situation, the frequency of the excited plasmons lies in the GHz
range, with the velocity $v_\mathrm{pl}$ roughly 100 times slower than the
speed of light $c$. The corresponding wave length $\lambda_\mathrm{pl} \sim
100~\mu$m then is much larger than the size $\sim 0.1~\mu$ of the quantum
point contact.  Assuming a coupling region $0.1~\mu\mathrm{m} < L <
\lambda_\mathrm{pl}$, we can ignore the $k$ dependence in $\hat I_k(t)$.

We assume that initially, at time $t=-t_0<0$, no plasmons are excited and the
state of the total system (QPC and bosonic modes) is described by the
factorized density matrix $\hat \rho(-t_0) = \hat \rho_\mathrm{pl}(-t_0)
\otimes \hat\rho_{\rm\scriptscriptstyle QPC}(-t_0)$. At a later time $t=0$,
the probability density $p_n(\Omega_1,\dots,\Omega_n)$ to find $n$ plasmons
with frequencies $\Omega_1, \dots, \Omega_n$ can be defined in terms of the
time-ordered evolution operator $\hat{S}(0,-t_0) =T \exp[-(i/\hbar)
\int_{-t_0}^0 \hat H_\mathrm{int}(t^\prime) dt^\prime ]$,
\begin{equation}
      p_n = \mbox{Tr} \bigl\{ \hat{\cal P}_n(\Omega_1,...,\Omega_n)
      \,\hat S(0,-t_0) \hat\rho(0) \hat S^\dagger(0,-t_0)\bigr\},
\end{equation}
where $\hat\rho(t)$ is the time dependent density matrix of the electronic
system (including scattering at the QPC, interactions between electrons,
etc.), $\hat{\cal P}_n = |k_1,\dots,k_n\rangle \langle k_1,\dots,k_n|$ is the
projection operator on the state with $n$ bosons of frequencies $\Omega_i=
v_\mathrm{pl} k_i$, and the trace is taken with respect to the total system (note
that $\hat{\cal P}_n(\vec \Omega) = \hat{\cal P}_n(\vec k) \prod \delta(\vec
\Omega- v_\mathrm{pl} \vec k$); the projection does not select specific
directions of the emitted plasmons). The Taylor expansion of the evolution
operator $\hat{S}(0,t)$ up to the lowest non-trivial order provides us with the
(one plasmon) probability density,
\begin{equation}
      p_1(\Omega) = \gamma^2\frac{\Omega}{\hbar v_\mathrm{pl}^3}
      \int\limits_{-t_0}^0 ds\,d\tau\, e^{i\Omega(\tau -s)}
      \, \langle \hat I(s) \hat I(\tau) \rangle,
      \label{P1}
\end{equation}
where $\langle\dots\rangle$ denotes the average over the electronic system.
The next term in the series generates the two-plasmon probability density
\begin{eqnarray}
      \label{P2}
      &&p_2(\Omega_1,\Omega_2) = \frac{\gamma^4}{4} \frac{\Omega_1}{\hbar
      v_\mathrm{pl}^3} \frac{\Omega_2}{\hbar v_\mathrm{pl}^3}
      \int\limits_{-t_0}^0 ds_1\,ds_2\, d\tau_1\,d\tau_2\, \\
      &&\nonumber
      \qquad
      \times \bigl\langle T_-\{ \hat I(s_1) \hat I(s_2)\}
      T_+\{ \hat I(\tau_1) \hat I(\tau_2)\} \bigr\rangle
      \\
      &&\qquad\times \bigl( e^{i\Omega_1(\tau_1-s_1)}
      e^{i\Omega_2(\tau_2-s_2)}+ e^{i\Omega_1(\tau_1-s_2)}
      e^{i\Omega_2(\tau_2-s_1)} \nonumber
      \\
      &&\qquad\qquad
       + \Omega_1 \leftrightarrow \Omega_2
      \bigr),
      \nonumber
\end{eqnarray}
where $T_+$ and $T_-$ are the time ordering operators in the forward and
backward directions, respectively. In the following, we assume $t_0\gg
\Omega^{-1}, \Omega_{1,2}^{-1}$ and regularize the time integrals in Eqs.\
(\ref{P1}) and (\ref{P2}) at the lower limit by introducing a small damping
factor $\exp(-\eta|\tau|)$ with $\eta \ll \Omega$.  Physically, $\eta^{-1}$
corresponds to the decay time of the plasmon excitations propagating in the
wave guide.

The above time integrals over current correlators can be expressed through the
spectral power of current fluctuations. Assuming a stationary situation, the
current correlators in Eqs.\ (\ref{P1}) and (\ref{P2}) depend only on relative
times; the second-order noise correlator $S^{(2)}$ then can be defined through
\begin{equation}
      S^{(2)}(\omega) = \int d\tau\, e^{-i\omega\tau} \,
      \langle\langle \hat I(\tau) \hat I(0) \rangle\rangle,
\end{equation}
and the fourth-order correlator $S^{(4)}$ reads
\begin{eqnarray}
      \nonumber
      &&S^{(4)}(\omega_1,\omega_2,\omega_3) = \int d\tau_1 d\tau_2
      d\tau_3\, e^{-i\omega_1\tau_1 -i\omega_2\tau_2
      -i\omega_3\tau_3}
      \\
      &&\qquad \times \, \langle\langle \hat I(\tau_1+\tau_2+\tau_3)
      \hat I(\tau_2+\tau_3) \hat I(\tau_3) \hat I(0) \rangle\rangle.
      \label{S4}
\end{eqnarray}
The noise correlators $S^{(2)}$ and $S^{(4)}$ involve {\it irreducible}
current correlators, while the expressions for $p_1(\Omega)$ and
$p_2(\Omega_1,\Omega_2)$ are expressed in terms of {\it reducible} quantities.
Expressing the reducible correlator through irreducible ones,
\begin{equation}
      \langle \hat I(s) \hat I(\tau)\rangle =
      \langle\langle \hat I(s) \hat I(\tau) \rangle\rangle +
      \langle \hat I(s)\rangle \langle \hat I(\tau) \rangle,
\end{equation}
and using the Fourier transform $\langle\langle \hat I(s) \hat I(\tau) \rangle
\rangle = \int (d\omega/2\pi)$ $\exp[i\omega (s-\tau)] S^{(2)}(\omega)$, we
can perform the time integrals in the expression Eq.\ (\ref{P1}) for
$p_1(\Omega)$. In the stationary regime, the average current $\langle
I(t)\rangle = \bar I$ through the QPC is independent of time and we obtain
the intermediate form
\begin{equation}
      p_1(\Omega) = \gamma^2 \frac\Omega{\hbar v_\mathrm{pl}^3}
      \biggl( \int\frac{d\omega}{2\pi} \frac{S^{(2)}(\omega)}{
      (\omega-\Omega)^2+\eta^2} + \frac{\bar{I}^2}{\Omega^2
      +\eta^2} \biggr).
\end{equation}
In the limit $\eta\ll \Omega$, we can approximate the Lorentzian by a
$\delta$-function, $\eta/[(\omega -\Omega)^2+\eta^2] \approx \pi
\delta(\omega-\Omega)$, and carry out the integral over $\omega$ to arrive at
the final result,
\begin{equation}
      p_1(\Omega) = \gamma^2 \frac\Omega{\hbar v_\mathrm{pl}^3}
      \Bigl(\frac{S^{(2)}(\Omega)}{2\eta}  +
      \frac{\bar{I}^2}{\Omega^2 +\eta^2} \Bigr).
      \label{P1o}
\end{equation}
Furthermore, the contribution from the irreducible part of the current
correlator to $p_1(\Omega)$ is a factor $\Omega/\eta\gg 1$ larger than the
contribution from the reducible part $\propto \bar{I}^2$, as follows from the
estimate $S^{(2)}(\Omega) \sim \bar{I}^2/\Omega$; we will drop the reducible
part in our further analysis below.

Expressing $p_2(\Omega_1,\Omega_2)$ through irreducible correlators is more
involved: below, we keep only those terms of the reducible fourth-order
current correlator which give non-vanishing contributions to $p_2$ in the
limit $\eta \ll \Omega$,
\begin{eqnarray}
      &&\langle \hat{I}(s_2) \hat {I}(s_1) \hat {I}(\tau_1)
      \hat{I}(\tau_2)\rangle = \langle \langle \hat{I}(s_2)
      \hat {I}(s_1) \hat {I}(\tau_1) \hat{I}(\tau_2)\rangle \rangle
      \nonumber\\
      &&\qquad + \langle\langle \hat{I}(s_2) \hat{I}(\tau_2)\rangle\rangle
      \langle\langle \hat{I}(s_1) \hat{I}(\tau_1) \rangle\rangle +
      \dots, \label{eq:4-th}
\end{eqnarray}
where we assume $\tau_1>\tau_2$, $s_1>s_2$ and $\dots$ denotes terms with
different time orderings as well as third-order cumulants providing irrelevant
contributions.  The probability distribution $p_2$ can then be expressed
through the noise correlators $S^{(4)}$ and $S^{(2)}$; the contribution from
the reducible part of the fourth-order current correlator reads
\begin{eqnarray}
      \nonumber
      &&p_2^{(2)}(\Omega_1,\Omega_2) = \gamma^4
      \frac{\Omega_1}{\hbar v_\mathrm{pl}^3}
      \frac{\Omega_2}{\hbar v_\mathrm{pl}^3}
      \frac{S^{(2)}(\Omega_1)
      S_2^{(2)}(\Omega_2)}{4\eta^2}
      \\
      &&\qquad\times \Bigl( 1 + \frac{\eta^2}{(\Omega_1-
      \Omega_2)^2+\eta^2}\Bigr).
      \label{P2r}
\end{eqnarray}
This term describes the excitation of two plasmons due to the uncorrelated
scattering of independent electrons.  For two bosons with equal frequency
$|\Omega_1-\Omega_2|\ll \eta$, this contribution is enhanced by a factor $2$ as
compared with the probability to emit two bosons with different frequencies
(see Eq.~(\ref{P1o})),
\begin{equation}
      p_2^{(2)}(\Omega_1,\Omega_2) =\left\{ \begin{array}{ll}
      p_1(\Omega_1) p_1(\Omega_2),& |\Omega_1-\Omega_2|\gg \eta,\\
      \noalign{\vspace{3pt}}
      2p_1^2(\Omega_1),& |\Omega_1-\Omega_2|\ll \eta.
      \end{array}
      \right.
\end{equation}
This enhancement is a quantum mechanical time-interference effect: for
$|\Omega_1-\Omega_2|\ll\eta$ we cannot know which boson was emitted
first during the measurement time $\eta^{-1}$ and the amplitudes of both
alternatives have to be added, resulting in a constructive interference
between them.

Next, we concentrate on the contribution $p_2^{(4)}$ arising from the
irreducible part of the fourth-order current correlator in Eq.\
(\ref{eq:4-th}). After integration over times, we arrive at the expression
\begin{widetext}
\begin{eqnarray}
      \nonumber
      p_2^{(4)}(\Omega_1,\Omega_2) &=& \gamma^4
      \frac{\Omega_1}{\hbar v_\mathrm{pl}^3}
      \frac{\Omega_2}{\hbar v_\mathrm{pl}^3}
      \int \frac{d\omega_1 d\omega_2
      d\omega_3}{(2\pi)^3}
      \frac{S^{(4)}(\omega_1,\omega_2,\omega_3)}{
      (\omega_2-\Omega_1-\Omega_2)^2+4\eta^2}
      \\
      &&\qquad \times
      \Bigl(
      \frac1{(\omega_1-\Omega_1-i\eta)(\omega_3-\Omega_1+i\eta)} +
      \frac1{(\omega_1-\Omega_1-i\eta)(\omega_3-\Omega_2+i\eta)} +
      \Omega_1\leftrightarrow \Omega_2 \Bigr).
      \label{P2i}
\end{eqnarray}

Combining Eqs.\ (\ref{P2r}) and (\ref{P2i}) and approximating the Lorentzian
in Eq.\ (\ref{P2i}) by a $\delta$-function, one finally arrives at the
expression for the probability density to emit two plasmons,
\begin{equation}
   \label{P2o}
      p_2(\Omega_1,\Omega_2) = \gamma^4 \frac{\Omega_1}{\hbar v_\mathrm{pl}^3}
      \frac{\Omega_2}{\hbar v_\mathrm{pl}^3}
      \frac{1}{4\eta} Q^{(4)}(\Omega_1,\Omega_2)
      + p_2^{(2)}(\Omega_1,\Omega_2)
\end{equation}
with
\begin{equation}
      \label{Q4}
      Q^{(4)}(\Omega_1,\Omega_2)=
      \int \frac{d\omega_1 d\omega_3}{(2\pi)^2}
      \Bigl[ \frac{S^{(4)}(\omega_1,\Omega_1\!+
      \!\Omega_2,\omega_3)}{(\omega_1\!-\!\Omega_1\!-\!i\eta)(\omega_3
      \!-\!\Omega_1\!+\!i\eta)} + \frac{S^{(4)}(\omega_1,\Omega_1\!+
      \!\Omega_2,\omega_3)}{(\omega_1\!-\!\Omega_1\!-\!i\eta)(\omega_3
      \!-\!\Omega_2\!+\!i\eta)} +\Omega_1 \leftrightarrow \Omega_2
      \Bigr].
\end{equation}
\end{widetext}

The resulting probability densities Eqs.\ (\ref{P1o}) and (\ref{P2o}) involve
second-order current correlators at positive frequencies and integrals over
fourth-order current correlators with integration kernels concentrated near
positive frequencies as well. This feature is easily understood: initially
there are no bosonic excitations, hence within our lowest-order calculation
the only processes contributing to $p_1(\Omega)$ and $p_2(\Omega_1,\Omega_2)$
are due to plasmon emission---reabsorbtion processes involving negative
frequency correlators show up only within a higher-order analysis.
\begin{figure}
   \includegraphics[width=5cm]{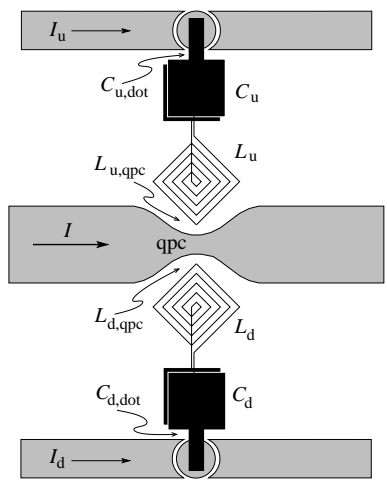}
   \caption[]
   {Setup to probe the emission of photons from a QPC into free space. The
   photons are picked up by the $LC$ circuits `up' and `down', which are
   inductively ($L_\mathrm{u(d), qpc}$) coupled to the QPC and capacitively
   ($C_\mathrm{u(d), dot}$) coupled to the quantum dots. The latter are tuned
   to minimal or maximal transmission and thus operate as quadratic detectors.
   Photons absorbed in the $LC$-circuits modulate the charge on the capacitors
   $C_\mathrm{u(d)}$ and, through coupling to the quantum dots, the probe
   currents $I_\mathrm{u}$ and $I_\mathrm{d}$.}
   \label{fig:setup_LC}

\end{figure}

Finally, we note that our results hold true, up to a numerical factor of order
unity, for the emission of (long wave length) photons as well; a corresponding
setup with $LC$ circuit pick-ups and quadratic detectors is sketched in
figure \ref{fig:setup_LC}.  For the photonic emission, one needs to replace the
plasmon wave velocity $v_\mathrm{pl}$ by the speed of light $c$ and set the
geometry factor $\gamma = 1$ in Eqs.\ (\ref{P1o}) and (\ref{P2o}). We will see
below, cf.\ Eq.\ \ref{eq:alpha}, that the basic dimensionless parameter
governing the numerical value of our results is $\alpha \propto (e^2/\hbar
v_\mathrm{pl}) (v_\mathrm{F}/v_\mathrm{pl})^2$. As a consequence, the one- and
two-photon emission rates are suppressed by the small factors
$(v_\mathrm{pl}/c)^3$ (with $v_\mathrm{pl}/c \sim 10^{-2}$ typically) and
$(v_\mathrm{pl}/c)^6$, respectively, when real three-dimensional photons are
emitted into space.  In the following, we use the terms plasmons and photons
synonymously.

\section{Photon counting statistics for a quantum current}
\label{sec:count}

It is a well known result due to Glauber~\cite{glauber_63} that a {\it
classical} current $I(t)$ produces a coherent state of the electromagnetic
field, with the width of the Poisson statistics of photo counts for each mode
given by the corresponding Fourier coefficient $I(\omega)$ of the current.  This
implies that photons are emitted independently, i.e., the joint probability to
emit two photons with frequencies $\Omega_1$ and $\Omega_2$ is in fact a
product of single photon probabilities, $P_2(\Omega_1,\Omega_2) =
P_1(\Omega_1) P_1(\Omega_2)$.

For a {\it quantum} current the fluctuations appear due to the scattering of
separate electrons. In the scattering process, the electrons experience an
acceleration and emit {\it Bremsstrahlung} radiation. The fermionic statistics
induces a correlated flow of the electrons incoming from the voltage biased
reservoir and separate electrons are scattered one by one. The photons emitted
in the scattering of separate electrons then inherit their fermionic
correlations.  On the other hand, one electron may produce several photons
during the scattering event, in which case these photons are bunched. The
deviation from poissonian statistics thus is a competition between these two
processes.

Using the results of the previous section, we can find the probability
$P_1(\Omega)= p_1(\Omega)\, d\Omega$ for the emission of plasmons in the
frequency interval $d\Omega$ around $\Omega$ and $P_2(\Omega_1,\Omega_2) =
p_2(\Omega_1,\Omega_2)\,d\Omega_1 d\Omega_2$ for the emission of two plasmons
at given frequencies within the non-overlapping frequency windows $d\Omega_1$
and $d\Omega_2$ around $\Omega_1,~\Omega_2$; for overlapping frequency
intervals $\Omega_1\approx\Omega_2=\Omega$, the determination of $P_2(\Omega)$
requires proper integration of the density $p_2(\Omega_1,\Omega_2)$.

While the results on the one- and two-plasmon processes provide us with only
limited information on the photo counting statistics, they are nevertheless
sufficient to tell us about important quantum signatures in the radiation. In
particular, the statistical properties of the emitted plasmons is conveniently
described by the irreducible probability correlator,
\begin{equation}
      \bar{P}_2(\Omega_1,\Omega_2)= P_2(\Omega_1,\Omega_2)
      -P_1(\Omega_1) P_1(\Omega_2).
      \label{PC}
\end{equation}
A positive sign of $\bar{P}_2(\Omega_1,\Omega_2)$ indicates that plasmons are
{\it bunching}, i.e., they are preferentially emitted simultaneously. In the
opposite case of negative correlations, $\bar{P}_2(\Omega_1,\Omega_2)<0$, the
plasmons are {\it anti-bunched}, implying that the probability to find the
second plasmon emitted right after the first is suppressed.

\subsection{Single plasmon emission}
\label{sec:single}

In the following, we assume that the emitted radiation is collected from a
finite region of length $L < \lambda_\mathrm{pl}$ and we average the
current operators in Eq.\ (\ref{Ik}) using a specific kernel of the form
$f(x)=\exp(-|x|/L)$. First, we concentrate on the single plasmon probability
$P_1(\Omega)$. Inserting the ($T=0$) second-order noise correlator\cite{S2}
\begin{equation}
      S_{x_1 x_2}^{(2)}(\Omega) = \frac{e^2}{\pi}\, TR\,
      e^{i(\Omega/v_{\rm\scriptscriptstyle F})[|x_2|-|x_1|]} 
      (\omega_0\!-\!\Omega)
      \Theta(\omega_0\!-\!\Omega)
\end{equation}
into Eq.~(\ref{P1o}) ($\Theta(x)$ is the Heavyside function), one arrives at
the result for the single plasmon emission probability in the form,
\begin{equation}
      P_1(\Omega) = \frac{2\alpha}{\pi}\, TR\,
      \Bigl(\frac{\omega_0\!-\!\Omega}{\Omega}\Bigr)\,
      \frac{\Theta(\omega_0\!-\!\Omega)}{1\!+\!(\Lambda/L)^2}
      \,\frac1\eta\,d\Omega,
      \label{P1S}
\end{equation}
with $\omega_0=eV/\hbar$ the voltage frequency, $\Lambda=v_\mathrm{F}/\Omega
\sim 100~\mu$m the characteristic spatial scale of the current fluctuations,
and $\alpha$ is the dimensionless parameter, 
\begin{equation}\label{eq:alpha}
      \alpha = \gamma^2\, \Bigl( \frac{e^2}{\hbar
      v_\mathrm{pl}} \Bigr) \Bigl( \frac{v_\mathrm{F}}{v_\mathrm{pl}}
      \Bigr)^2 \sim \frac{\gamma^2}{100}.
\end{equation}
Note that with $d$ and $w$ of order 1 $\mu$m \cite{burke}, hence the geometry
factor $\gamma$ is of order unity. On the other hand, the parameter $\alpha$
is reduced due to the residual impedance mismatch between the quantum point
contact and the waveguides, cf.\ Fig.\ \ref{fig:setup}.

The probability $P_1(\Omega)$ describes the Bremsstrahlung plasmon emission
due to electron scattering at the barrier. Although one may expect the
probability $P_1(\Omega)$ to increase for a more effective scatterer with
$T\rightarrow 0$, the result is in fact proportional to $T(1-T)$ and vanishes
in the tunneling limit. This peculiarity of the Bremsstrahlung appears due to
the Fermi statistics: the electron coming in from the biased reservoir has to
relax to a state with a lower energy after the plasmon emission. Since all
electron scattering states of the biased reservoir with lower energy are
filled, the only possibility for the electron to decay is into an empty
scattering state of the unbiased reservoir. This process requires tunneling of
the electron through the barrier and hence the probability $P_1(\Omega)$ turns
out proportional to $T$. The sharp suppression of $P_1(\Omega)$ when the
plasmon energy $\hbar\Omega$ is larger than the applied bias $eV$ can be
explained in the same way: an electron cannot find an empty state to emit such
a `high-energy' plasmon.

\subsection{Correlated plasmon emission}
\label{sec:correlated}

Next, we analyze the two-plasmon probability $P_2(\Omega_1,\Omega_2)$, see
Eq.~(\ref{P2o}), which involves irreducible fourth-order noise correlators.
Below, we consider the case of an {\it extended} interaction region, where the
signal is collected over a region of size $L$ larger than the characteristic
spacial scale $\Lambda = v_\mathrm{F}/\Omega_{1,2}\sim 100~\mu$m of current
fluctuations, where we consider typical Fermi velocities $v_\mathrm{F}\sim
10^7$cm/s in GaAs heterostructures and frequencies in the GHz regime. The
calculation of the expression $Q^{(4)}(\Omega_1,\Omega_2)$ in Eq.\ (\ref{Q4})
is carried out in the Appendix, cf.\ Eq.\ (\ref{S4o}) for the final result;
the probability $P_2(\Omega_1,\Omega_2)$ for two-plasmon emission in
non-overlapping frequency intervals then reads
\begin{widetext}
\begin{eqnarray} \label{P2F}
      P_2(\Omega_1,\Omega_2) &=& \frac{4\alpha^2}{\pi\eta}\,
      \bigl( RT(T\!-\!R)^2
      \Theta\bigl(\omega_0-\Omega_\Sigma\bigr)
      \bigl(\omega_0-\Omega_\Sigma\bigr)
      -2(RT)^2\Theta\bigl(\omega_0-\Omega_>\bigr)
      \bigl(\omega_0-\Omega_>\bigr)\bigr)
      \frac{d\Omega_1}{\Omega_1}
      \frac{d\Omega_2}{\Omega_2}
      \\
      &&\quad
      +\frac{4\alpha^2}{(\pi\eta)^2}\, (RT)^2 \Theta\bigl(
      \omega_0  -\Omega_1\bigr)\Theta\bigl( \omega_0 -\Omega_2\bigr)
      \bigl( \omega_0 -\Omega_1\bigr) \bigl( \omega_0
      -\Omega_2\bigr) \Bigl[1
      +\frac{\eta^2}{(\Omega_1-\Omega_2)^2+\eta^2} \Bigr]
      \frac{d\Omega_1}{\Omega_1}\frac{d\Omega_2}{\Omega_2},
      \nonumber
\end{eqnarray}
%
where $\Omega_\Sigma=\Omega_1+\Omega_2$ and $\Omega_> = \max\{\Omega_1,
\Omega_2\}$. The first term $\propto 1/\eta$ in the above expression
originates from the irreducible fourth-order noise correlator $Q^{(4)}$, while
the second term $\propto 1/\eta^2$ is a contribution from the reducible part
$p^{(2)}_2$ of the fourth-order current correlator, see Eq.~(\ref{P2o}).

For non-overlapping frequency bands with $|\Omega_1-\Omega_2| \gg \eta$, the
second term in the bracket $[\dots]$ in the reducible part is small and what
remains is equal to the product $P_1(\Omega_1)P_1(\Omega_2)$ of single plasmon
probabilities. Hence, the irreducible probability correlator $\bar{P}_2$,
Eq.~(\ref{PC}) involves only the first irreducible term in Eq.~(\ref{P2F}),
\begin{equation}
      \bar{P}_2(\Omega_1,\Omega_2) = \frac{4\alpha^2}{\pi\eta}\,
      \Bigl[ RT(T\!-\!R)^2\Theta\bigl(\omega_0-\Omega_\Sigma\bigr)
      \bigl(\omega_0-\Omega_\Sigma\bigr)
      -2(RT)^2\Theta\bigl(\omega_0-\Omega_>\bigr)
      \bigl(\omega_0-\Omega_>\bigr)\Bigr]
      \frac{d\Omega_1d\Omega_2}{\Omega_1\Omega_2}.
      \label{PCi}
\end{equation}
\end{widetext}

In the opposite situation where the frequency bands overlap, the
density $p_2(\Omega_1,\Omega_2)$ has to be properly integrated over
frequencies; the second term (of Lorentzian shape) in $p_2^{(2)}$,
cf.\ Eq.\ (\ref{P2r}), then provides a contribution $\pi/\eta$.
For a narrow frequency band $\delta \Omega/\Omega \ll 1$ far from the
voltage frequency, $|\omega_0-\Omega|\sim\Omega$, this reducible
contribution dominates over the irreducible one originating from
$Q^{(4)}$ and the irreducible probability correlator Eq.~(\ref{PC})
is dominated by the reducible part of the fourth-order current
correlator,
\begin{equation}
      \bar{P}_2(\Omega) \approx \frac{4\alpha^2}{\pi\eta} (TR)^2
      \Theta(\omega_0-\Omega) (\omega_0-\Omega) \frac{\omega_0-\Omega}{\Omega}
      \frac{\delta\Omega}{\Omega}.
\end{equation}

The situation changes if one detects plasmon in the overlapping frequency band
reaching the voltage frequency, $\Omega_1, \Omega_2 \in [
\omega_0-\Delta\Omega, \omega_0]$, $\Delta\Omega <\omega_0/2$. In this
situation one needs to take the remaining integrals over the frequencies
$\Omega_1, \Omega_2$ in Eq.~(\ref{P2F}) exactly. The result takes the form
\begin{equation}
      P_2(\Omega) = - \frac{8\alpha^2}{3\pi}\, (RT)^2 \,
      \frac{(\Delta\Omega)^3}{\eta\Omega^2} +\frac{4\alpha^2}{3\pi}\,
      (RT)^2 \frac{(\Delta\Omega)^3}{\eta\Omega^2},
\end{equation}
where the first term originates from the irreducible part of the fourth-order
current correlator, while the second term has its origin in the reducible
contribution. We then find that the irreducible contribution dominates in this
regime.

In summary, the  irreducible correlator $\bar{P}_2$ at different frequencies
$\Omega_1 \neq \Omega_2$ involves a competition between bunching (first term
in (\ref{PCi})) and anti-bunching (second term in (\ref{PCi})). When the
frequency intervals strongly overlap we encounter two regimes: 1) detecting
the plasmon in a narrow frequency band far from the voltage frequency,
$|\omega_0-\Omega|\sim\Omega\gg \delta\Omega$, the reducible contribution of
the current correlator always dominates thus resulting in bunched plasmon
radiation; 2) measuring the plasmon in the frequency band
$[\omega_0-\Delta\Omega, \omega_0]$, $\Delta\Omega<\omega_0/2$, near the
voltage frequency, the irreducible contribution of the current correlator for
a single-channel conductor is two times larger then the reducible one in
absolute values, resulting in anti-bunched radiation.

It is the irreducible correlator $\bar{P}_2$ at different frequencies
$\Omega_1 \neq \Omega_2$ which is the most interesting quantity measuring the
non-trivial correlator $Q^{(4)}$. Thus, in order to study the fourth-order
electron noise effects on the plasmon statistics, one needs to be able to
distinguish between the frequencies of the emitted plasmon during the
measurement time $\eta^{-1}$.

We thus concentrate on the detection of plasmon with distinguishable
frequencies. The probability correlator $\bar{P}_2(\Omega_1,\Omega_2)$, see
Eq.~(\ref{PCi}), involves two terms with opposite signs, a positive one
proportional to $\Theta(\omega_0-\Omega_1-\Omega_2)$ and a negative
contribution $\propto \Theta(\omega_0-\max\{\Omega_1,\Omega_2\})$. Applying a
small voltage bias $\omega_0<\max\{\Omega_1,\Omega_2\}$, the scattered
electrons do not have the possibility to emit two plasmons with frequencies
$\Omega_1$ and $\Omega_2$ due to the restriction of the Fermi statistics on
the final electron state and hence $\bar{P}_2(\Omega_1,\Omega_2)=0$.

In the intermediate voltage regime $\max\{\Omega_1,\Omega_2\} <\omega_0<
\Omega_1+\Omega_2$, a single electron can emit only one plasmon, either of
frequency $\Omega_1$ or $\Omega_2$. Thus the corresponding contribution
$\propto \Theta(\omega_0 -\max\{\Omega_1, \Omega_2\})$ is due to the
scattering of different (fermionically correlated) electrons. At zero
temperature, these electrons arrive at the scatterer in perfect order with a
time separation $\tau_V\sim h/eV$. Hence, the second plasmon is emitted only
after the time $\tau_V$, resulting in an {\it anti-bunched} radiation
statistics and a negative probability correlator $\bar{P}_2 (\Omega_1,
\Omega_2) < 0$. When a single electron creates only one plasmon, the complexity
of the emission process is reduced considerably, allowing for the
determination of the full counting statistics of the emitted
radiation~\cite{beenakker_01,beenakker_04}.

Increasing the voltage beyond $\hbar(\Omega_1+\Omega_2)$, the above single
plasmon production is augmented by processes where one electron emits two
plasmons in a single scattering event; this process generates {\it bunched}
radiation and hence the corresponding term $\propto \Theta(\omega_0 -\Omega_1
-\Omega_2)$ is positive, see Eq.~(\ref{PCi}). The overall sign of the
probability correlator $\bar{P}_2(\Omega_1, \Omega_2)$ then appears in a
competition between the two processes creating plasmons individually or in
pairs. Changing the parameters of the quantum point contact, one can control
the relative weight of the two contributions. For example, at $T=1/2$ the
two-plasmon process $\propto TR(T-R)^2$ vanishes while the single-plasmon term
$\propto 2(TR)^2$ is maximal, resulting in  anti-bunched radiation in the
whole two-plasmon voltage regime $\Omega_1 + \Omega_2 <\omega_0< \Omega_1+
\Omega_2+ \min\{\Omega_1, \Omega_2\}$ (we consider three-plasmon processes
involving at least one plasmon with frequency $\Omega_1$ and one with
$\Omega_2$). Alternatively, in the tunneling limit $T\ll 1$, the emission of
bunched plasmons is the dominant process at $\omega_0>\Omega_1+ \Omega_2$. A
further increase of the bias voltage may lead to multi-plasmon production
processes, where more than two plasmons are created by a single scattered
electron.  However, such processes appear only in higher orders of
perturbation theory, while we have restricted ourselves to the fourth order in
the coupling constant.

Finally, we analyze the case of a multi-channel conductor in the diffusive
regime. The probability correlator $\bar{P}_2(\Omega_1, \Omega_2)$ for
non-overlapping frequencies in the multi-channel case can be obtained by
summing the terms in Eq.~(\ref{PCi}) over the channel index $n$ with
appropriate values for the scattering coefficients $T_n$ and $R_n$. The
distribution of the transmission eigenvalues $T$ in the diffusive conductor is
given by the Dorokhov distribution function~\cite{dorokhov},
\begin{equation}
      \rho(T)=\frac{G}{2G_0}\,\frac1{T\sqrt{1-T}},
\end{equation}
where $G$ is the average conductance and $G_0=2e^2/h$.  Averaging Eq.\
(\ref{PCi}) over $T$, the probability correlator $\bar{P}_2(\Omega_1,
\Omega_2)$ for the radiation emitted from the diffusive conductor is given by
the expression (we assume $\Omega_1 \neq \Omega_2$),
\begin{eqnarray}
      &&\bar{P}_2(\Omega_1,\Omega_2) =
      \frac{4\alpha^2}{\pi\eta}\,\frac{G}{G_0}
      \Bigl( \frac{11}{105}\,\Theta\bigl(\omega_0-\Omega_\Sigma\bigr)
      \bigl(\omega_0-\Omega_\Sigma\bigr)
      \nonumber\\
      && -\frac{4}{35}\,\Theta\bigl(\omega_0-\Omega_>\bigr)
      \bigl(\omega_0-\Omega_>\bigr)\Bigr)
      \frac{d\Omega_1}{\Omega_1}
      \frac{d\Omega_2}{\Omega_2}.
      \label{PCid}
\end{eqnarray}
Quite remarkably, the result is negative, indicating anti-bunched plasmons,
even in the two-plasmon regime $\omega_0 > \Omega_1 +\Omega_2$. This provides,
besides the measurements of the average current and noise, another consistency
test for the Dorokhov distribution function.

\subsection{Measurement setup}
\label{sec:measurement}

Next, we relate the probabilities $P_1(\Omega)$ and $P_2(\Omega_1,\Omega_2)$
to physically measurable quantities. In a realistic experiment, see
Refs.~\onlinecite{gabelli_04,portier}, a two-terminal quantum point contact is
realized in a two-dimensional electron gas inserted between two transmission
lines, cf.\ Fig.\ \ref{fig:setup}. The excited plasmon excitations in the 2DEG
induce an $ac$-electric current signal in the two (left and right)
transmission lines. The transmitted signals are independently amplified by two
cryogenic amplifiers and then passed through frequency filters (selecting
proper frequencies $\Omega_{1,2}$) followed by quadratic detectors. Thus the
signal taken at the end of each transmission line is proportional to the power
emitted from the QPC,
\begin{equation}
      \hat{W}_i = A_i  \hbar\Omega_i \hat{n}(\Omega_i) d\Omega_i
      + w_i,\quad i=1,2,
\end{equation}
where $A_i$ is the amplification factor, $\hat{n}(\Omega) d\Omega$ is the
plasmon production rate, i.e., the number of excited plasmons per unit time
within the frequency band $d\Omega$, and $w_i$ is a parasitic power due to the
intrinsic noise of the amplifiers and the vacuum fluctuations of the bosonic
mode. Given the life time (or detection time) $1/\eta$ of the plasmons, cf.\
Eqs.~(\ref{P1o}) and~(\ref{P2o}), the plasmon production rate $\hat{n}(\Omega)
d\Omega$ relates to the plasmon occupation number $\hat{N}(\Omega)$ via
$\hat{n}(\Omega) d\Omega = \eta \hat{N}(\Omega)$.

Since the noise signals $w_{1,2}$ are not correlated for the
different transmission lines, $\langle \delta w_1 \delta w_2\rangle
= 0$, $\delta w_i = w_i -\langle w_i\rangle$, the irreducible
cross-correlator between the two transmission lines, $Q_{12}=
\langle \hat{W}_1\hat{W}_2\rangle - \langle \hat{W}_1\rangle\langle
\hat{W}_2\rangle$, involves only the irreducible cross-correlator of
the plasmon occupation numbers emitted into each line,
\begin{equation}
      Q_{12} =A_1 A_2\, \hbar\Omega_1 \hbar\Omega_2\,
      \eta^2\,\langle\langle\hat{N}(\Omega_1) \hat N(\Omega_2)
      \rangle\rangle.
      \label{WW}
\end{equation}
Next, we express the plasmon number correlator $\langle\langle \hat{N}
(\Omega_1) \hat N(\Omega_2) \rangle\rangle$ through the probabilities
$P_1(\Omega)$ and $P_2(\Omega_1,\Omega_2)$ calculated above. Up to
fourth-order in perturbation theory, we have
\begin{eqnarray}
      &&\langle \hat{N}(\Omega)\rangle \approx p_1(\Omega)d\Omega
      + \Bigl[\int\!\!d\Omega^\prime p_2(\Omega,\Omega^\prime)\Bigr]\, d\Omega,
      \\
      &&\langle \hat{N}(\Omega_1) \hat{N}(\Omega_2) \rangle
      \approx p_2(\Omega_1,\Omega_2) d\Omega_1  d\Omega_2,
\end{eqnarray}
where the second term in $\langle \hat{N}(\Omega)\rangle$ is a higher order
correction.  As a result, the irreducible cross correlator for the plasmon
number then assumes the simple form,
\begin{equation}
       \langle\langle \hat{N}(\Omega_1) \hat N(\Omega_2)
       \rangle\rangle= P_2(\Omega_1,\Omega_2)-P_1(\Omega_1)
       P_1(\Omega_2),
       \label{NN}
\end{equation}
involving only the probabilities $P_{2}(\Omega_1,\Omega_2)$ and $P_1(\Omega)$;
other terms are of higher order in the coupling constant.  The cross
correlator of the emitted power between two transmission lines then is
proportional to the probability correlator $\bar{P}_2(\Omega_1,\Omega_2)$, see
Eq.~(\ref{PC}), and the sign of the power cross correlator directly
characterizes the statistics of the emitted radiation.

Finally, we connect our results with the results of Beenakker and
Schomerus~\cite{beenakker_01,beenakker_04}, which are based on the
Glauber formula for photon counting. Within this approach the
relevant physical quantity to observe deviations from the poissonian
statistics is the variance of the detected particles: $\mbox{Var}(N)
= \langle N^2\rangle -\langle N\rangle^2$. Within lowest order
perturbation theory, $\mbox{Var}(N)$ can be expressed through the
probabilities $P_1$ and $P_2$,
\begin{eqnarray}
      \mbox{Var}(N) &=& 4P_2+P_1 -P_1^2
      \\
      &=& \langle N\rangle + 2P_2-P_1^2.
\end{eqnarray}
The sign of the combination $2P_2-P_1^2$ quantifies the deviation of the
photon statistics from the poissonian result $\mbox{Var}(N) = \langle
N\rangle$. Let us first concentrate on the regime where photons are measured
in a narrow frequency band $[\Omega -\delta\Omega/2, \Omega+\delta\Omega/2]$,
$\delta\Omega\ll \Omega$, far from the voltage frequency $|\omega_0- \Omega|
\sim \Omega$.  According to the results of Ref.~\onlinecite{beenakker_01} the
quantity $2P_2-P_1^2$ is proportional to the measurement time $\tau=1/\eta$,
i.e., terms $\propto 1/\eta^2$ mutually cancel between the terms $2P_2$ and
$P_1^2$ (the literal correspondence between our probabilities and those of
Beenakker and Schomerus is obtained by the substitution $\gamma_0\rightarrow
4\alpha/\Omega$).  Substituting our probabilities for this regime, we find
that, in contrast to the result of the Ref.~\onlinecite{beenakker_01}, the
leading contribution to $2P_2-P_1^2$ is proportional to $\eta^{-2}$,
\begin{eqnarray}
      2P_2-P_1^2 \sim \frac{4\alpha^2}{\pi^2} \bigl[ \delta\Omega
      \tau \bigr]^2\, (RT)^2
      \Bigl(\frac{\omega_0-\Omega}\Omega\Bigr)^2.
\end{eqnarray}

Technically this difference arises from the fact, that $P_1^2$ cannot
compensate for the contribution from the reducible current correlator in
$2P_2$, see Eq.~(\ref{P2F}). The physical reason for the observed distinction
between our results and the results of Ref.~\onlinecite{beenakker_01} lies in
the different measurement procedure. The Glauber photodetection procedure
implies a real {\it counting} of the photons, i.e., in addition to the number
of photons one gains extra information on the time of the detection. In
contrast, in our detection scheme we do not {\it count} the plasmons---our
probabilities $P_1(\Omega)$ and $P_2(\Omega_1,\Omega_2)$ contain only
information about the total number of the plasmons at the end of the
measurement. Thus, we do not know which plasmon with frequency $\Omega_1$ or
$\Omega_2$ was detected first. As a result, our probability to observe two
plasmons $P_2(\Omega)$ is two times larger than that obtained via the Glauber
detection scheme, and hence no compensation occurs in our case. Instead, it is
the quantity $\bar{P}_2 = P_2 - P_1^2$ which exhibits the proper cancellation
and provides a measure for the non-trivial correlations in the plasmon
statistics in our analysis. The same situations occurs in the regime where
plasmons are detected in the frequency band $[\omega_0-\Delta\Omega,
\omega_0]$, $\Delta\Omega< \omega_0/2$; our probability $P_2(\Omega)$ is again
twice larger than the result in Ref.~\onlinecite{beenakker_04}.

\section{Measurement of the fourth order current correlators}
\label{sec:current}

As mentioned in the introduction, the measurement of the plasmon statistics
also reveals information on the current fluctuations in the QPC. In
particular, the probabilities $P_1(\Omega)$ and $P_2(\Omega_1,\Omega_2)$
provide information on the second- and fourth-order noise correlators at
finite frequencies.  Furthermore, the measurement of high-order current
correlators, or alternatively, high-order transmitted charge cumulants, is a
non trivial issue. It turns out, that the measurement of the irreducible
correlator $\bar{P}_2$ provides direct access to the fourth-order charge
correlator.

The charge statistics is conveniently analyzed in a Gedanken experiment, where
the transmitted charge is measured with the help of a spin-$1/2$ counter
\cite{levitov_96,comment}. Using the expression $\hat Q_t = \int_0^t \hat
I(\tau) d\tau$ for the transmitted charge, the corresponding generating
function involves the specific time ordering
\begin{equation}
      \chi (\lambda) = 
      \langle T_- [\exp(i\lambda\hat Q_t /2)]
              T_+ [\exp(i\lambda\hat Q_t /2)] \rangle,
      \label{chicl}
\end{equation}
where $T_-$ and $T_+$ denote backward and forward time-ordering operators. The
resulting statistics turns out to be binomial with the proper electron charge
describing the transmitted carriers\cite{levitov_96}. The fourth-order (zero
frequency) charge cumulant is given by a weighted combination of various time
orderings, 
\begin{eqnarray}
      &&\langle\langle \hat Q_t^4 \rangle\rangle =
      \frac1{16} \int_0^t d\tau_1  d\tau_2 d\tau_3 d\tau_4 \, 
      \bigl[ \langle\langle 
      T_-\bigl(\hat I_{\tau_1} \hat I_{\tau_2} \hat I_{\tau_3}\hat I_{\tau_4}
      \bigr) \rangle\rangle \nonumber\\
      &&+4\langle\langle 
      T_-\bigl(\hat I_{\tau_1}\hat I_{\tau_2}\hat I_{\tau_3}\bigr)
      \hat I_{\tau_4} \rangle\rangle 
      + 6 \langle\langle 
      T_-\bigl(\hat I_{\tau_1}\hat I_{\tau_2}\bigr)T_+\bigl(\hat I_{\tau_3} 
      \hat I_{\tau_4}\bigr) \rangle\rangle
      \nonumber\\
      &&+4 \langle\langle 
      \hat I_{\tau_1}T_+\bigl(\hat I_{\tau_2} \hat I_{\tau_3}\hat I_{\tau_4}
      \bigr)\rangle\rangle 
      + \langle\langle 
      T_+\bigl(\hat I_{\tau_1}\hat I_{\tau_2}\hat I_{\tau_3}\hat I_{\tau_4}
      \bigr) \rangle\rangle\bigr],
      \label{I4c}
\end{eqnarray}
and the result assumes the form
\begin{equation}
      \langle\langle \hat Q_t^4 \rangle\rangle
      =2 e^4T(1-T)(6T^2-6T+1)\frac{eVt}{h},
      \label{Q4c}
\end{equation}
with the poissonian limit restored in the tunneling regime.

While it is clear that this (binomial) result manifests itself when the charge
transport is analyzed with a spin-1/2 counter, the question can be posed
whether a more realistic experiment, e.g., the present plasmon detection
experiment, can be used to measure this result.  

Indeed, the measurement of the probability correlator $\bar{P}_2(\Omega_1,
\Omega_2)$ tests the time-ordered fourth-order current correlator, reminding
about the spin-1/2 detection scheme of Ref.~\onlinecite{levitov_96}, cf.\
Eqs.~(\ref{P2}) and (\ref{I4c}). Assuming an extended measurement where the
emitted radiation is collected from a region near the QPC with a size $L$ 
larger than the characteristic length $\Lambda$, the explicit
calculation of the frequency integrals in Eq.~(\ref{P2o}) gives a result for
$\bar{P}_2(\Omega_1,\Omega_2)$ at low frequencies $\Omega_1,\Omega_2\ll
eV/\hbar$ and $|\Omega_1-\Omega_2|\gg \eta$, see Eq.~(\ref{PCi}), which
coincides with the fourth-order charge correlator in Eq.\ (\ref{Q4c}),
\begin{equation}
      \bar{P}_2(\Omega_1,\Omega_2) =\alpha^2 T(1-T)(6T^2\!-6T+1)
      \frac{eV}{h\eta}\, \frac{d\Omega_1 d\Omega_2}{\Omega_1
      \Omega_2}.
\end{equation}
We thus conclude that for an extended measurement scheme with $L\gg\Lambda$,
the probability correlator $\bar{P}_2(\Omega_1, \Omega_2)$ at low frequencies
is proportional to the fourth-order charge cumulant $\langle\langle
\hat{Q}^4_t\rangle\rangle$ with $t=\eta^{-1}$. 

\section{Conclusion}\label{sec:conclusion}

We have presented a perturbative calculation of the statistics of plasmon
emission from electrons scattered at a quantum point contact.  In our
analysis, we determine the probability densities $p_n$ to find $n$ plasmons
with prescribed frequencies $\Omega_1, \dots, \Omega_n$ during a measuring
time $t_0 \sim 1/\eta$; our perturbative calculation includes terms up to
fourth order in the interaction Hamiltonian and allows us to calculate one-
($p_1$) and two- ($p_2$) plasmon processes.  These probability densities are
related to second- and fourth- order current correlators and hence measuring
the plasmon statistics provides also information on higher order current
correlators.

Our central quantity is the ireducible probability correlator
$\bar{P}_2(\Omega_1, \Omega_2) = P_2(\Omega_1,\Omega_2) - P_1(\Omega_1)
P_1(\Omega_2)$, which we find to provide the most valuable information if
measured at different frequencies $|\Omega_1-\Omega_2| \gg \eta$.  Its sign
provides information on the statistics of plasmon emission, which arises from a
competition between bunching due to for multi-plasmon emission from one
electron and anti-bunching imprinted onto the plasmons by the regular stream of
incident electrons. The character of the emitted radiation can be tuned
between bunched and anti-bunched by changing the voltage $V$ and the
transmission $T$ across the QPC. Measuring the power cross-correlator in
different transmission lines as done in recent experiments
\cite{gabelli_04,portier} provides experimental access to this quantity.
Within the usual photo detection scheme instead
\cite{beenakker_01,beenakker_04}, the role of $\bar{P}_2 = P_2-P_1^2$ is
played by the deviation of the variance from the Poisson value, $\mbox{Var}(N)
- \langle N\rangle = 2 P_2 - P_1^2$. The discrepancy in the factor 2 in front
of $P_2$ is a consequence of the different measurement techniques, providing
more detailed information in the photo detection scheme. This rather innocent
looking difference in fact requires the definition of a different measurement
quantity for the two cases of `counting' (set of single photon measurements)
and `collecting' (single projection of a plasmon number state at the end); the
former requires the analysis of $\mbox{Var}(N) - \langle N\rangle$, while the
latter forces one to discuss $\bar{P}_2$.  Finally, we have shown that the
irreducible correlator $\bar{P}_2(\Omega_1,\Omega_2)$ coincides (up to a scale
factor) with the fourth-order charge cumulant $\langle\langle \hat Q_t^4
\rangle\rangle$ and hence provides practical access to this quantity.

We thank Fabien Portier and Klaus Ensslin for discussions and acknowledge
financial support by the Swiss National Science Foundation, the CTS-ETHZ, and
the Russian Foundation for Basic Research under grant No.\ 08-02-00767-a.

\appendix \section{Fourth-order current correlator} \label{cucor}

Using the scattering matrix approach, we first calculate the irreducible
fourth-order current correlator in the time representation,
\begin{equation}
      C(\vec{x},\vec{t}\>) = \langle\langle \hat I(x_1,t_1)
      \hat I(x_2,t_2) \hat I(x_3,t_3) \hat I(x_4,t_4)
      \rangle\rangle,
\end{equation}
and then determine the expression $Q^{(4)}(\Omega_1,\Omega_2)$ in
Eq.~(\ref{Q4}).

We assume a scattering process described by energy independent scattering
amplitudes $r$, $\bar{r}$, and $t$ within a region near the QPC located at the
origin $x=0$; here, $r$ and $\bar{r}$ denote reflection amplitudes for
electrons coming from the left and right reservoirs, and $t$ is the
transmission amplitude.  Linearizing the energy-momentum dispersion relation
near the Fermi level, the electron current operator takes the form,
\begin{widetext}
\begin{eqnarray}
      &&\hat I(x>0,t) = \frac{e}{h} \int\!\! d\epsilon
      d\epsilon^\prime
      \bigl( T\, \hat{a}_{\epsilon}^\dagger
      \hat{a}_{\epsilon^\prime} + rt^*\, \hat{a}_{\epsilon}^\dagger
      \hat{b}_{\epsilon^\prime} +r^*t\, \hat{b}_{\epsilon}^\dagger
      \hat{a}_{\epsilon^\prime} +R\,\hat{b}_{\epsilon}^\dagger
      \hat{b}_{\epsilon^\prime}\bigr) e^{i(\epsilon-
      \epsilon^\prime)(t-\frac{x}{v_{\rm\scriptscriptstyle F}})/\hbar} 
         - \hat{b}_{\epsilon}^\dagger
      \hat{b}_{\epsilon^\prime} e^{i(\epsilon-
      \epsilon^\prime)(t+\frac{x}{v_{\rm\scriptscriptstyle F}})/\hbar},
      \\
      &&\hat I(x<0,t) = -\frac{e}{h} \int\!\! d\epsilon
      d\epsilon^\prime
      \bigl( R\, \hat{a}_{\epsilon}^\dagger
      \hat{a}_{\epsilon^\prime} + \bar{r}^*t\, \hat{a}_{\epsilon}^\dagger
      \hat{b}_{\epsilon^\prime} +\bar{r}t^*\, \hat{b}_{\epsilon}^\dagger
      \hat{a}_{\epsilon^\prime} +T\,\hat{b}_{\epsilon}^\dagger
      \hat{b}_{\epsilon^\prime}\bigr) e^{i(\epsilon-
      \epsilon^\prime)(t+\frac{x}{v_{\rm\scriptscriptstyle F}})/\hbar} 
            + \hat{a}_{\epsilon}^\dagger
      \hat{a}_{\epsilon^\prime} e^{i(\epsilon-
      \epsilon^\prime)(t-\frac{x}{v_{\rm\scriptscriptstyle F}})/\hbar},
\end{eqnarray}
\end{widetext}
where $\hat{a}_\epsilon$ and $\hat{b}_\epsilon$ are annihilation operators for
electron scattering states incoming from the left and right reservoirs,
respectively (we assume spinless electrons; $T=|t|^2$ and $R=1-T$ are
transmission and reflection probabilities, $v_{\rm\scriptscriptstyle F}$ is
the Fermi velocity).

The current operator $\hat I(x,t)$ can be written as a sum $\hat I(x,t)=\hat
I^{\scriptscriptstyle -}(\xi^-) + \hat I^{\scriptscriptstyle +}(\xi^+)$ of
outgoing and incoming currents $I^{\scriptscriptstyle -}(\xi^-)$ and
$I^{\scriptscriptstyle +}(\xi^+)$ which depend only on the retarded variables
$\xi^\pm=t\pm|x|/v_{\rm\scriptscriptstyle F}$.  Below, we concentrate on the
current fluctuations to the right of the barrier.  Introducing an additional
index $\alpha = \pm $ denoting the incoming and outgoing currents, we rewrite
the current operators in a compact form,
\begin{equation}
      \hat{I}^\alpha(\xi^\alpha)=\frac{e}{h}
      \int d\epsilon d\epsilon^\prime
      e^{i(\epsilon-\epsilon^\prime)\xi^\alpha/\hbar}
      \sum\limits_{ij} \hat c_i^\dagger(\epsilon)\,
      A^\alpha_{ij}\,\hat c_j(\epsilon^\prime),
\end{equation}
where we have defined $\hat c_1(\epsilon) = \hat a_\epsilon$, $\hat
c_2(\epsilon) = \hat b_\epsilon$, and the $2\times2$ matrices
$A^{\scriptscriptstyle\pm}$
\begin{equation}
      A^{\scriptscriptstyle +}=
      \left(\begin{array}{cc}
      T& t^*r\\  r^*t& R
      \end{array}\right), \qquad
      A^{\scriptscriptstyle -}=
      \left(\begin{array}{cc}
      0&0\\0&-1
      \end{array} \right)
\end{equation}
defined to the right of the barrier. In order to calculate the fourth-order
current correlator $C(\vec{x},\vec{t}\>)$, we have to average over all
possible products of four current operators, 
\begin{equation}
      C^{\vec\alpha}(\vec{\xi}\,) = \langle\langle
      \hat{I}^{\alpha_1}(\xi_1^{\alpha_1})
      \hat{I}^{\alpha_2}(\xi_2^{\alpha_2})
      \hat{I}^{\alpha_3}(\xi_3^{\alpha_3})
      \hat{I}^{\alpha_4}(\xi_4^{\alpha_4})
      \rangle\rangle,
\end{equation}
with $C(\vec{x},\vec{t}\>)=\sum_{\vec\alpha} C^{\vec\alpha}(\vec{\xi}\,)$.
Below, we use the shorthand $\xi_i^{\alpha_i}=\xi_i$ and $A^{\alpha_i}=A_i$
and put $\hbar=1$. Using Wick's theorem and taking averages over the
reservoirs, we arrive at the expression 
\begin{widetext}
\begin{eqnarray}
      \nonumber
      &&C^{\vec\alpha}(\vec\xi\,) = \frac{e^4}{(2\pi)^4}
      \int\prod_{i=1}^4 d\epsilon_i \bigl(
      e^{i\epsilon_1(\xi_1-\xi_2)}
      e^{i\epsilon_2(\xi_2-\xi_3)}
      e^{i\epsilon_3(\xi_3-\xi_4)}
      e^{i\epsilon_4(\xi_4-\xi_1)}
      \mbox{Tr}\{ N(\epsilon_1) A_1\bar{N}(\epsilon_4)
      A_4N(\epsilon_3)A_3 N(\epsilon_2) A_2\}
      \\
      \nonumber
      &&\qquad\qquad+
      e^{i\epsilon_1(\xi_1-\xi_4)}
      e^{i\epsilon_2(\xi_2-\xi_1)}
      e^{i\epsilon_3(\xi_3-\xi_2)}
      e^{i\epsilon_4(\xi_4-\xi_3)}
      \mbox{Tr}\{ N(\epsilon_1) A_1\bar{N}(\epsilon_2)
      A_2\bar{N}(\epsilon_3)A_3 \bar{N}(\epsilon_4) A_4\}
      \\
      \nonumber
      &&\qquad\qquad-
      e^{i\epsilon_1(\xi_1-\xi_4)}
      e^{i\epsilon_2(\xi_2-\xi_3)}
      e^{i\epsilon_3(\xi_3-\xi_1)}
      e^{i\epsilon_4(\xi_4-\xi_2)}
      \mbox{Tr}\{ N(\epsilon_1) A_1\bar{N}(\epsilon_3)
      A_3 N(\epsilon_2)A_2 \bar{N}(\epsilon_4) A_4\}
      \\
      \nonumber
      &&\qquad\qquad-
      e^{i\epsilon_1(\xi_1-\xi_3)}
      e^{i\epsilon_2(\xi_2-\xi_1)}
      e^{i\epsilon_3(\xi_3-\xi_4)}
      e^{i\epsilon_4(\xi_4-\xi_2)}
      \mbox{Tr}\{ N(\epsilon_1) A_1\bar{N}(\epsilon_2)
      A_2 \bar{N}(\epsilon_4)A_4 N(\epsilon_3) A_3\}
      \\
      \nonumber
      &&\qquad\qquad-
      e^{i\epsilon_1(\xi_1-\xi_3)}
      e^{i\epsilon_2(\xi_2-\xi_4)}
      e^{i\epsilon_3(\xi_3-\xi_2)}
      e^{i\epsilon_4(\xi_4-\xi_1)}
      \mbox{Tr}\{ N(\epsilon_1) A_1\bar{N}(\epsilon_4)
      A_4 N(\epsilon_2)A_2 \bar{N}(\epsilon_3) A_3\}
      \\
      &&\qquad\qquad-
      e^{i\epsilon_1(\xi_1-\xi_2)}
      e^{i\epsilon_2(\xi_2-\xi_4)}
      e^{i\epsilon_3(\xi_3-\xi_1)}
      e^{i\epsilon_4(\xi_4-\xi_3)}
      \mbox{Tr}\{ N(\epsilon_1) A_1\bar{N}(\epsilon_3)
      A_3 \bar{N}(\epsilon_4)A_4 N(\epsilon_2) A_2\} \bigr),
      \label{C4}
\end{eqnarray}
with the $2\times 2$ matrices $N(\epsilon)$ and $\bar{N}(\epsilon)$
defined as
\begin{equation}
      N(\epsilon) = \left( \begin{array}{cc} n_L(\epsilon)&0
      \\0&n_R(\epsilon) \end{array}\right),
      \quad \bar{N}(\epsilon) = \mathbbm{1} - N(\epsilon),
\end{equation}
and $n_L(\epsilon)$ and $n_R(\epsilon)$ the Fermi distribution
functions of the left and right electronic reservoirs. Next, we
integrate over the energies in Eq.~(\ref{C4}) using 
\begin{eqnarray}
      \int d\epsilon\, (1\!-\!n_{L/R}(\epsilon)) e^{i\epsilon\xi} =
      \frac{i\pi\theta\, e^{\pm i\omega_0\xi/2}}{\sinh[\pi\theta(\xi
      +i\delta^\prime)]},
      \qquad
      \int d\epsilon\, n_{L/R}(\epsilon) e^{i\epsilon\xi} =
      \frac{-i\pi\theta\, e^{\pm i\omega_0\xi/2}}{\sinh[\pi\theta(\xi
      -i\delta^{\prime\prime})]},
\end{eqnarray}
where $\theta$ is the temperature of the fermionic reservoirs and
$\omega_0=eV/\hbar$ is the voltage frequency defined by the bias voltage $V$
applied to the QPC. $\delta^\prime, \delta^{\prime\prime}>0$ are
regularization parameters; for an energy independent transparency, they are of
order $\delta^\prime \sim \hbar/(E_c-E_{\rm\scriptscriptstyle F})$ and
$\delta^{\prime\prime} \sim \hbar/E_{\rm\scriptscriptstyle F}$, where $E_c$ is
the energy width of the conduction band and $E_{\rm\scriptscriptstyle F}$ the
Fermi energy. Below, we define a single regularization parameter $\delta =
\max\{\delta^\prime,\delta^{\prime\prime}\}$ and assume the zero temperature
limit $\theta=0$. After integration and using these simplifications we obtain
the expression
\begin{eqnarray}
      \nonumber
      &&C^{\vec\alpha}(\vec\xi\,) = \frac{e^4}{(2\pi)^4} \Bigl(
      -\frac{\mbox{Tr}\{p(\xi_1-\xi_4)A_1 p(\xi_2-\xi_1) A_2
      p(\xi_3-\xi_2) A_3 p(\xi_4-\xi_3)A_4\}+c.c.}{(\xi_1-\xi_4-i\delta)
      (\xi_2-\xi_1+i\delta)(\xi_3-\xi_2+i\delta)(\xi_4-\xi_3+i\delta)}
      \\
      \nonumber
      &&\qquad\qquad\qquad\quad
      + \frac{\mbox{Tr}\{p(\xi_1-\xi_4)A_1 p(\xi_3-\xi_1)
      A_3 p(\xi_2-\xi_3) A_2 p(\xi_4-\xi_2)A_4\}+c.c.}{(\xi_1-\xi_4-i\delta)
      (\xi_3-\xi_1+i\delta)(\xi_3-\xi_2+i\delta)(\xi_4-\xi_2+i\delta)}
      \\
      &&\qquad\qquad\qquad\quad
      + \frac{\mbox{Tr}\{p(\xi_1-\xi_3)A_1 p(\xi_2-\xi_1)
      A_2 p(\xi_4-\xi_2) A_4 p(\xi_3-\xi_4)A_3\}+c.c.}{(\xi_1-\xi_3-i\delta)
      (\xi_2-\xi_1+i\delta)(\xi_4-\xi_2+i\delta)(\xi_4-\xi_3+i\delta)}
      \Bigr),
\end{eqnarray}
\end{widetext}
where $p(\xi)$ is the $2\times2$ diagonal matrix,
\begin{equation}
      p(\xi) = \left( \begin{array}{cc} e^{i\omega_0\xi}&0\\0&1
      \end{array}\right).
\end{equation}

Next, we consider the irreducible contribution to the probability
$P_2(\Omega_1,\Omega_2)$, see Eq.~(\ref{P2o}).  This contribution is
proportional to the frequency integral over the fourth-order noise correlator
$S^{(4)}(\omega_1, \Omega_1 + \Omega_2, \omega_2)$ with a specific kernel.
Assuming a stationary situation (i.e., only relative times are relevant) and
changing from frequency to time variables in Eq.\ (\ref{Q4}), we have to
calculate the expression
\begin{eqnarray}
      &&Q^{(4)}(\Omega_1,\Omega_2) =
      \int\limits_{-\infty}^0 d\tau_1\,
      \int\limits_0^\infty d\tau_2\,
      \tilde{C}(\tau_1,\Omega_1\!+\!\Omega_2,\tau_2)
      \nonumber\\
      &&
      \bigl( e^{-i\Omega_1(\tau_1+\tau_2)} +
      e^{-i(\Omega_1\tau_1+\Omega_2\tau_2)} + \Omega_1
      \leftrightarrow \Omega_2 \bigr),
      \label{tav}
\end{eqnarray}
where we have defined
\begin{equation}
      \tilde{C}(\tau_1,\Omega,
      \tau_2) = \int d\tau\,
      \tilde{C}(\tau_1,\tau,
      \tau_2)\, e^{-i\Omega\tau}.
      \label{C4O}
\end{equation}
Here, $\tilde{C}(\tau_1,\tau_2,\tau_3)$ is the coordinate averaged
correlator $C(\vec{x},\vec{t}\>)$ expressed in terms of the relative
time variables,
\begin{eqnarray}
      \tilde{C}(\tau_1,\tau_2,\tau_3) &=& \!\int\! d^4 x F(\vec{x}\,)\\
      \nonumber
      &&\quad \times 
      C(\vec{x};\tau_1\!+\!\tau_2\!+ \!\tau_3,\tau_2\!+\!\tau_3,\tau_3,0),
\end{eqnarray}
with the kernel $F(\vec{x}\,) = \prod_i f(x_i)$ describing the spatial
extension of the plasmon.

Next, we find the non-vanishing contributions $\tilde{C}^{\vec\alpha}
(\tau_1,\Omega >0,\tau_2)$ to the correlator $\tilde{C}(\tau_1,\Omega,\tau_2)
=\sum_{\vec\alpha} \tilde{C}^{\vec\alpha}(\tau_1,\Omega,\tau_2)$ defined in
Eq.~(\ref{C4O});  these are identified as those with $\vec\alpha \in\{(----),
(+---), (-+--), (--+-), (---+), (-++-), (+-+-), (-+-+),(+--+)\}$.  The
corresponding correlators can be written in the form,
\begin{widetext}
\begin{equation}
      \tilde{C}^{\vec\alpha}(\tau_1,\Omega,\tau_2)=
      \frac{e^4}{4\pi^3}\,e^{i\Omega(\tau_1+\tau_2)/2}
      \!\int\! d^4 x\, F(\vec{x}\,)\, I_\Omega^{\vec\alpha}(z_1,z_2)
      \exp\Bigl(\frac{i\Omega(\alpha_1|x_1|
      \!+\!\alpha_2|x_2|\!-\!\alpha_3|x_3|\!-\!\alpha_4|x_4|)}
      {2v_{\rm\scriptscriptstyle F}}
      \Bigr),
\end{equation}
where we have introduced the new retarded variables, 
\begin{equation}
      z_1(\alpha_1,\alpha_2)
      =\tau_1+\frac{\alpha_1|x_1|-\alpha_2|x_2|}{v_{\rm\scriptscriptstyle F}}, 
       \qquad
      z_2(\alpha_3,\alpha_4)
      =\tau_2+\frac{\alpha_3|x_3|-\alpha_4|x_4|}{v_{\rm\scriptscriptstyle F}},
\end{equation}
and the functions $I_\Omega^{\vec\alpha}(z_1,z_2)$ have the form,
\begin{eqnarray}
      \nonumber
      && I^{----}_\Omega =  2RT(T\!-\!R)^2
      \Theta(\omega_0\!-\!\Omega)\Bigl( \cos[{\omega_0}(z_1\!-\!z_2)/2]
      \frac{\sin[(\omega_0-\Omega)(z_1\!+\!z_2)/2]}{(z_1\!+\!z_2)z_1z_2}
      +  z_2\rightarrow -z_2 \Bigr)
      \\ \nonumber
      && \qquad\qquad+
      2(RT)^2\, \Theta(\omega_0\!-\!\Omega/2)\Bigl(
      \frac{\sin[(\omega_0-\Omega/2)(z_1\!+\!z_2)]}{(z_1\!+\!z_2)z_1z_2} +
      z_2\rightarrow -z_2 \Bigr),
      \\ \nonumber
      && I^{+---}_\Omega =RT
      \Theta(\omega_0\!-\!\Omega) e^{-i\omega_0z_1/2}\Biggl[
      \Bigl( (Re^{-i\omega_0z_2/2}
      +Te^{i\omega_0z_2/2}) \frac{\sin[(\omega_0-\Omega)(z_1\!+\!z_2)/2]}
      {(z_1\!+\!z_2)(z_1\!-\!i\delta)z_2} + z_2\rightarrow-z_2\Bigr)
      \\ \nonumber
      &&\qquad\qquad +2R\cos[\omega_0z_2/2]\Bigr(
      \frac{\sin[(\omega_0-\Omega)(z_1\!-\!z_2)/2]}{(z_1\!-\!z_2)z_1z_2}
      +z_2\rightarrow -z_2 \Bigr)\Biggr],
\end{eqnarray}
\begin{eqnarray}
      \nonumber
      && I^{-++-}_\Omega = RT\Theta(\omega_0\!-\!\Omega)
      e^{i\omega_0(z_1+z_2)/2} \Biggl[ \Bigl( \frac{\sin[(\omega_0
      -\Omega)(z_1\!+\!z_2)/2]}{(z_1\!+\!z_2)z_1z_2} +z_2\rightarrow -z_2\Bigr)
      -\frac{\sin[(\omega_0-\Omega)(z_1+z_2)/2]}{(z_1\!+\!z_2)(z_1\!-\!i\delta)
      (z_2\!-\!i\delta)} \Biggr],
      \\ \nonumber
      && I^{+-+-}_\Omega = RT\Theta(\omega_0\!-\!\Omega)
      e^{i\omega_0(z_2-z_1)/2} \Biggl[ \Bigl( \frac{\sin[(\omega_0
      -\Omega)(z_1\!+\!z_2)/2]}{(z_1\!+\!z_2)z_1z_2} +z_2\rightarrow -z_2\Bigr)
      +\frac{\sin[(\omega_0-\Omega)(z_1-z_2)/2]}
      {(z_1\!-\!z_2)(z_1\!-\!i\delta)(z_2\!-\!i\delta)} \Biggr].
\end{eqnarray}
\end{widetext}
The remaining functions $I^{\vec\alpha}_\Omega$ can be expressed
through the four above,
\begin{eqnarray}
      \nonumber
      &&I^{-+--}_\Omega(z_1,z_2,\delta) =
      I^{+---}_\Omega(-z_1,z_2,-\delta),
      \\
      \nonumber
      &&I^{--+-}_\Omega(z_1,z_2,\delta) =
      I^{+---}_\Omega(-z_2,z_1,-\delta),
      \\
      &&I^{---+}_\Omega(z_1,z_2,\delta) =
      I^{+---}_\Omega(z_2,z_1,\delta),
      \\
      \nonumber
      &&I^{-+-+}_\Omega(z_1,z_2,\delta) =
      I^{+-+-}_\Omega(-z_1,-z_2,-\delta),
      \\
      \nonumber
      &&I^{+--+}_\Omega(z_1,z_2,\delta)=
      I^{-++-}_\Omega(-z_1,-z_2,-\delta).
\end{eqnarray}

Finally, we have to perform the time integrals over $\tau_1$ and $\tau_2$ in
Eq.~(\ref{tav});, we regularize the divergent denominators in
$I^{\vec\alpha}_\Omega(z_1, z_2)$ using the Sokhotsky formula,
\begin{equation}
      \lim_{\delta\rightarrow 0^+}\frac1{z\pm i\delta} = {\cal
      P}\frac1z \mp i\pi \delta^\prime(z),
\end{equation}
where the $\delta$-function has to be understood as $\pi\delta^\prime(z) =
\delta/(z^2+\delta^2)$ with a finite width $\delta$.  The correlator
$Q^{(4)}(\Omega_1,\Omega_2)$ then takes the form
\begin{widetext}
\begin{eqnarray}
      &&Q^{(4)}(\Omega_1,\Omega_2) = \frac{e^4}{2\pi} RT
      \Theta(\omega_0\!-\!\Omega_1\!-\!\Omega_2)
      (\omega_0\!-\!\Omega_1\!-\!\Omega_2) \int d^4{x}\, F(\vec{x}\,)\,
      \Bigl\{
      (T-R)^2 e^{-i\frac{\Omega_1}{v_{\rm\scriptscriptstyle F}}(|x_1|-|x_4|)}
      e^{-i\frac{\Omega_2}{v_{\rm\scriptscriptstyle F}}(|x_2|-|x_3|)}
      \nonumber\\
      &&+\frac{T\!-\!R}2\Bigl( e^{i\frac{\Omega_1}{v_{\rm\scriptscriptstyle F}}
          (|x_1|+|x_4|)}
      e^{-i\frac{\Omega_2}{v_{\rm\scriptscriptstyle F}}(|x_2|-|x_3|)}
          g(|x_1|\!+\!|x_2|)+
      e^{-i\frac{\Omega_1}{v_{\rm\scriptscriptstyle F}}(|x_1|+|x_4|)}
      e^{-i\frac{\Omega_2}{v_{\rm\scriptscriptstyle F}}(|x_2|-|x_3|)}i
          g(|x_3|\!+\!|x_4|) +
      \Omega_1 \leftrightarrow \Omega_2 \Bigr)
      \nonumber\\
      &&+\frac14\,g(|x_1|\!+\!|x_2|)g(|x_3|\!+\!|x_4|) \Bigl(
      e^{-i\frac{\Omega_1}{v_{\rm\scriptscriptstyle F}}(|x_1|-|x_4|)}
      e^{i\frac{\Omega_2}{v_{\rm\scriptscriptstyle F}}(|x_2|-|x_3|)}+
      e^{-i\frac{\Omega_1}{v_{\rm\scriptscriptstyle F}}(|x_1|+|x_3|)}
      e^{i\frac{\Omega_2}{v_{\rm\scriptscriptstyle F}}(|x_2|+|x_4|)}
      +\Omega_1 \leftrightarrow \Omega_2 \Bigr)\Bigr\}
      \nonumber\\
      &&-2\frac{e^4}{2\pi} (RT)^2 \Theta(\omega_0-\max\{\Omega_1,\Omega_2\})
      (\omega_0-\max\{\Omega_1,\Omega_2\})
      \int d^4{x}\, F(\vec{x}\,)\, e^{-i\frac{\Omega_1}
            {v_{\rm\scriptscriptstyle F}}(|x_1|-|x_4|)}
      e^{-i\frac{\Omega_2}{v_{\rm\scriptscriptstyle F}}(|x_2|-|x_3|)},
      \label{S4o}
\end{eqnarray}
\end{widetext}
where $g(x)=1-\Theta_\delta(x)+\Theta_\delta(-x)$, with $\Theta_\delta(x)
=\int^x dy\, \delta^\prime(y)$ a Heaviside-like function with a finite width
$\lambda=v_{\rm\scriptscriptstyle F} \delta$ defined by the regularization
parameter $\delta$; with the parameters $\delta^\prime \sim
\hbar/(E_c-E_{\rm\scriptscriptstyle F})$ and $\delta^{\prime\prime} \sim
\hbar/E_{\rm\scriptscriptstyle F}$, we obtain $\lambda \sim \lambda_{\rm
\scriptscriptstyle F}$ of order of the Fermi wave length. Then $g(|x|)=1$ for
$|x|\ll \lambda$ and $g$ vanishes in the opposite case.

For a large collection area with $\lambda < \Lambda < L$, we can
drop all terms containing a factor $g(x)$ in Eq.\ (\ref{S4o}); the
integration over $\vec x$ generates a factor $\sim (v_{\rm
\scriptscriptstyle F}/\Omega)^4$ and we arrive at the result Eq.\
(\ref{P2F}) (we approximate the factors $1/(1+\Lambda^2/L^2) \approx
1$).

Taking into account the spin $1/2$ of the electron, the above expression has
to be multiplied by a factor $2$. Similarly, in a multi-channel situation, we
have to sum the correlators $Q_n^{(4)}(\Omega_1,\Omega_2)$ for all
channels $n$ with transparency $T_n$.

\end{document}